\documentclass[reprint,a4paper]{revtex4-1}

\usepackage[utf8]{inputenc}
\usepackage[french]{babel}
\usepackage{amsmath,soul}
\usepackage{amsfonts}
\usepackage{amssymb}
\usepackage[pdftex]{color,graphicx}

\begin{document}

\author{J. Pedregosa-Gutierrez}
\affiliation{Aix Marseille Univ, CNRS, PIIM, Marseille, France}
\email{jofre.pedregosa@univ-amu.fr}

\author{M. Mukherjee}
\affiliation{MajuLab, International Joint Research Unit UMI 3654, CNRS, Université Côte d’Azur, Sorbonne Université, National University of Singapore, Nanyang Technological University, Singapore}
\affiliation{Centre for Quantum Technologies, National University Singapore, Singapore, 117543, Singapore} 

\title{Realisation of homogeneous ion chain using surface traps.}

\date{\today}

\begin{abstract}
In a Radio-Frequency linear ion trap, 1D ion chains are routinely generated in laboratories around the world. They present a non-homogenous ion density along the chain. The possibility of generating uniformly distributed ion chain, where the distance between any adjacent ions is a constant, would open up new type of experiments in the context of Quantum Information and in the study of the Homogeneous Kibble-Zurek mechanism.

\end{abstract}


\maketitle
\section{Introduction}
Radio-frequency (RF) ion traps are used in many different contexts including quantum computing~\cite{bruzewicz_trapped-ion_2019}, cold reactions~\cite{PhysRevLett.97.243005}, cooling highly charged ions~\cite{RN174}, mass spectrometry~\cite{march_practical_1995} and quantum simulations~\cite{bruzewicz_trapped-ion_2019}.

In the context of quantum computing and simulation, novel scalable ion trap architecture, in particular the surface ion trap approach \cite{pino_demonstration_2021} has gained significant traction in recent years. Unlike a 3D linear ion trap geometry, the surface ion traps are relatively easy to micro-fabricate and hence the electrodes can be designed in an arbitrary shape and size~\cite{hucul_transport_2008, amini_toward_2010, mokhberi_optimised_2017}. However, implementing a large scale quantum algorithms require more than just a scalable ion trap; it is necessary a develop a scalable computing architecture capable of controlling $100$-$1000$s of qubits. Among such architectures, a quantum charged coupled device (QCCD)~\cite{RN134} has recently being realised in a rudimentary application~\cite{pino_demonstration_2021}. This requires the ions to be shuttled around while keeping alive the quantum state. This invariably requires full temporal and spatial control over the ion chain, thus requiring a large number of DC electrodes. Separation of two ions in different trapping regions~\cite{warschburger_experimental_2014}, transport of ions around the corners~\cite{hucul_transport_2008} are some of the examples challenges that needs careful consideration. While this approach may be viable for $100$s of qubits, it is often challenged by heating of the motional degree of freedom, an essential for two qubit gate operations. An alternative approach is based on distributed computing architecture, where each computing node either implements a QCCD architecture or uses multiple addressing beams for multiple ions without the need to shuttle the ions around~\cite{RN135}. The nodes are then connected via optical quantum communication channels. Considering the later option, the ions are not only addressed individually but also photons are collected from individual ions to create remote entanglement. Therefore, ion-ion separation is an important parameter to consider in distributed computing scheme. 

In case of a 1D chain of ion trapped and laser cooled in a harmonic potential, the ions are distributed non-uniformly. The ion-ion distance increases from the trap centre~\cite{kamsap_experimental_2017} to the edges. Technically, such a configuration is not convenient to address the ions using multiple laser beams. Furthermore, an uniform chain of trapped and laser cooled ions are test-beds for simulating a variety of quantum many body systems like the homogeneous Kibble-Zureck mechanism~\cite{RN161}, adiabatic entanglement generation in a Bose-Hubbard model~\cite{RN73} $\textit{etc}$. Adiabatic entanglement generation using the Bose-Hubbard model is easier to implement when the hopping probability is uniform over the ion chain, thus requiring uniform distribution of the ions. On the contrary the homogeneous Kibble-Zureck mechanism demands an uniform density of the ions in the chain as it is a strong assumption in the model.   

It was first suggested by Lin et al~\cite{lin_large-scale_2009} that the use of a quartic rather than a harmonic potential generates a near uniform ion distribution. More recently, Xie et al~\cite{xie_creating_2017} developed a feedback method to achieve a high degree of homogeneity on a surface trap experiment. Despite these developments, designing electrodes to generate uniform chain of ion with uniformity below $10\%$, remains a challenge. Often, the electric field generated by a non-regular shaped electrodes do not have an analytical solution, making it harder to calculate the required voltages and frequencies to operate such a surface ion trap. Most often, one relies on specialised software based on the Finite Element Method~(FEM) or the Boundary Element Methods~(BEM), to solve the Laplace/Poisson equations. An alternative numerical approach assumes that there is no gap between the electrodes: the gap-less(GL) approximation~\cite{house_analytic_2008}, which reflects the reality in case the electrodes are much longer than the real gaps.

Here, we report an alternative approach for designing multi-electrode surface ion trap in order to achieve an homogeneous ion chain. The homogeneity of the chain is aimed at reducing the qubit addressability error from individual laser beams and more importantly for experiments where the uniform density of strongly coupled oscillators are required such as in the Kibble-Zureck mechanism. The later puts stringent condition on the required uniformity for which the currently available techniques fail. 

The article is organised as follows: starting with a brief introduction to the surface ion trap architecture, the novel algorithm to calculate the voltages needed to generate a desired ion chain distribution is presented; It is then used to explore the possibility to obtain the desired topology in function of several experimental constraints and finally, molecular dynamics simulations are used to verify the degree of homogeneity of the ion-chain for various optimal trap parameters as obtained from the novel algorithm. 

\section{Basics of Surface Ion Traps}
\begin{figure}
\center
\scalebox{0.15}{\includegraphics{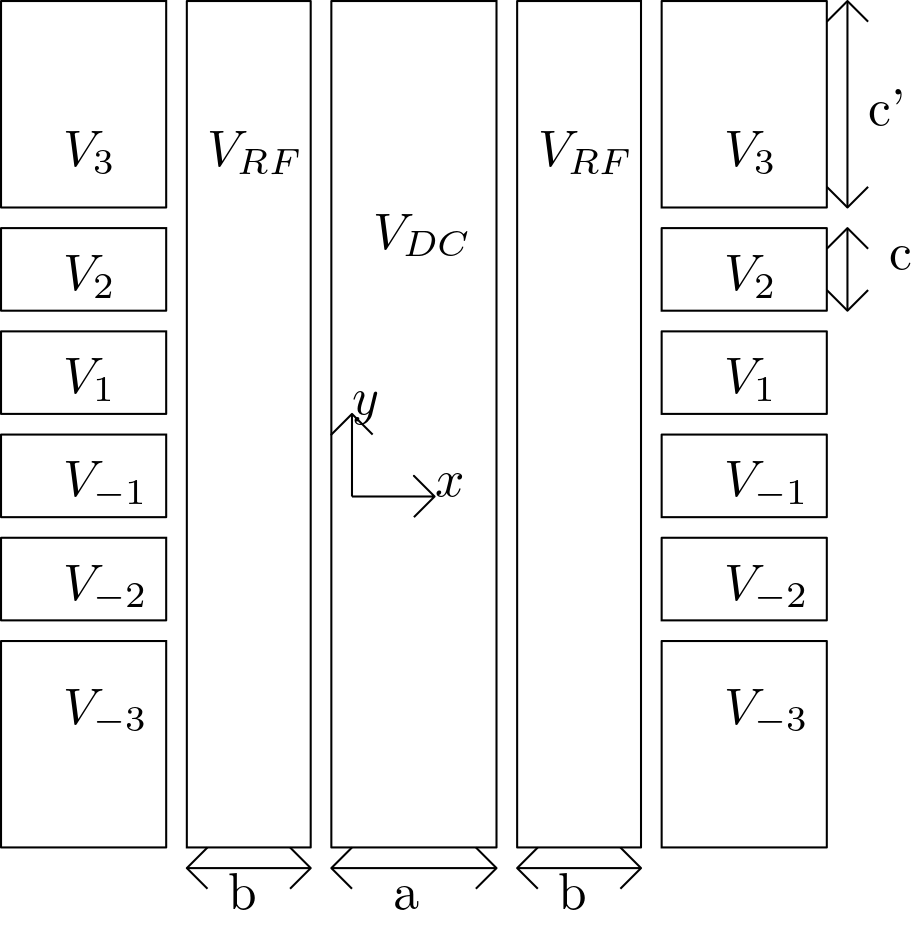}}
\caption{Example of a basic surface trap design with six pairs of control electrodes. For the ease of fabrication, it is usual to have the control electrodes of equal length except the outer ones which are larger. The surface trap electrodes are assumed to be in the $xz$ plane.}
\label{fig:designs}
\end{figure}

A Radio-Frequency~(RF) linear Paul trap uses a time dependent harmonic potential in three dimensions to generate a time averaged pseudo-potential to confine charge particles in 3D. The stability of the ion on the trap depends on the charge-to-mass ratio of the particle as well as the voltage and frequency of the applied time dependent field. A surface ion trap generates a similar pseudo-potential by planarizing the 3D electrodes to a plane. In this case, see figure~\ref{fig:designs}, the ions are trapped above the surface at a given height, $h$, corresponding to the location of the minimum of the pseudo-potential, $V_\text{PS}(\vec{r})$ at $\vec{r}$. The $V_\text{PS}(r)$ is given by:
\begin{eqnarray}\label{eq:Vps}
    V_{PS}(\vec{r}) = \frac{Q^{2}}{4 m \Omega^{2}}|\vec{E}_{RF}(\vec{r})|^{2}
\end{eqnarray}
where $Q$ and $m$ are the charge and the mass of the trapped particles, and $\vec{E}_\text{RF}(\vec{r})$ is the electric field generated by the RF electrodes at $\vec{r}$.

The methodology to calculate the electric field due to a planar infinite gap-less electrode has been discussed previously in~\cite{house_analytic_2008,schmied_electrostatics_2010}. Essentially, it consists of performing the following surface integral for each surface:
\begin{eqnarray}\label{eq:general_integral}
\Phi(x,y,z) = \int \frac{\phi(x',y')|z|}{2\pi ( (x-x')^2 +  (y-y')^2 +z^2  ) ^{3/2}} dx' dy'
\end{eqnarray}
with the surface potential $\phi(x,y) = \Phi(x,y,0)$ known.

For the particular case of a square surface, equation~\ref{eq:general_integral} can be solve analytically~\cite{house_analytic_2008}.

Such analytical results are used to compute important parameters, like the height at which the ions will be trapped, $h$. The parameter $h$ plays in fact an important role due to the strong dependency of the heating effects on $h$: $\propto h^4$ for the anomalous heating~\cite{turchette_heating_2000} and $\propto h^2$ for Johnson noise~\cite{wineland_experimental_1998}.

In the context of the GL approximation and for the case of equal width of the two RF electrodes (which is already assumed in figure~\ref{fig:designs}), the value of $h$ can be obtained analytically~\cite{house_analytic_2008}: $h = \sqrt{ab + a^2}$. Notice that the expression of $h$ depend only on geometric considerations and not on the applied voltages.

The trap depth can also be obtained analytically for this particular case~\cite{house_analytic_2008}:
\begin{equation}
V_\text{depth} = \frac{Q^2 V_{RF}^2}{\pi^2 m \Omega^2} \left( \frac{b}{(a+b)^2 + (a+b)\sqrt{2ab+a^2}} \right)^2
\end{equation}
where $V_\text{RF}$ and $\Omega$ are the voltage amplitude and the frequency of the applied voltage on the RF electrode.

Regarding the axial direction, the potential is defined by the control electrodes. Their total number, $n$, their width $c$, and the voltages applied on them, $V_{cc}$, will essentially determine the potential along the trap axis. In the following, we will present a methodology to obtain the $V_\text{cc}$ that will best approximate the desired potential.

\section{Optimization problem}
The objective is, given a number of ions, a desired ion-ion distance, $d$ and ion height, $h$, to find a set of parameters, $n$, $c$ and $V_{cc}$ that best approximated the desired experimental situation: that an equally distributed ion chain represents the equilibrium topology.

We start by defining the position of the ions as a uniform chain at $y=0$, $z=h$ and ion-ion distance of $d$. By imposing that the total force along all directions are equal to zero, we arrive at the following system of equations:

\begin{align}
\sum_{k=1}^{n}{ V_k E^{CC,x}_k(r_i) } = - E^{RF,x}_k(r_i) - E^{DC,x}_k(r_i) \\
\sum_{k=1}^{n}{ V_k E^{CC,y}_k(r_i) } = C_i -E^{RF,y}_k(r_i) - E^{DC,y}_k(r_i) \\
\sum_{k=1}^{n}{ V_k E^{CC,z}_k(r_i) } = - E^{RF,z}_k(r_i) - E^{DC,z}_k(r_i)
\end{align}

where $r_i = (x_i,y_i,z_i)$, $C_i = k_c Q \sum_j{ \frac{\vec{r}_{ij}}{|r_{ij}|^3}}$, $k_c$ is the Coulomb's constant and $E^{CC,x}_k$ ($E^{CC,y}_k$, $E^{CC,z}_k$) denotes the electric field generated by the control electrode $k$ in the $x$ ($y$, $z$) direction when 1V is applied to it. Similar for the RF and DC electrodes.

The above equations can be re-written as a linear least-squares problem~\cite{bevington_data_2003}.
\begin{eqnarray}\label{eq:leasst_square}
    \chi^2 = || A x - b ||^2
\end{eqnarray}
where $A$ is the design matrix and $b$ is the target vector.

In order to find the $V_i$ values that minimise $\chi^2 $, we have used the optimization solver found on the Scipy package~\cite{virtanen_scipy_2020} which allows us to impose bounds on the solutions. While the unbounded linear least-square problems lead to a unique solution based on matrix inversion, the bounded algorithms are iterative. The Scipy solver implements the Trust Region Reflective algorithm~\cite{branch_subspace_1999}, which iteratively solves trust-region sub-problems based on the on the different variables and the Bounded-Variable least-squares method~\cite{stark_bounded-variable_1995} where the variables are categorised in free or active, then the algorithm solves the case with the active variables fixed and the free ones as unbounded, effectively solving successively the unconstrained least-square problem. The presence of bounds is a requirement as the voltages supplies or the dielectric strength of the material used, will impose limitations on the maximum allowed voltages.

In the following, we will assume that only a half of the electrodes are operating independently. The other half is paired up. Following the notation of figure\ref{fig:designs}, it means that $V_i = V_{-i}$.

\begin{figure}
\center
\scalebox{0.35}{\includegraphics{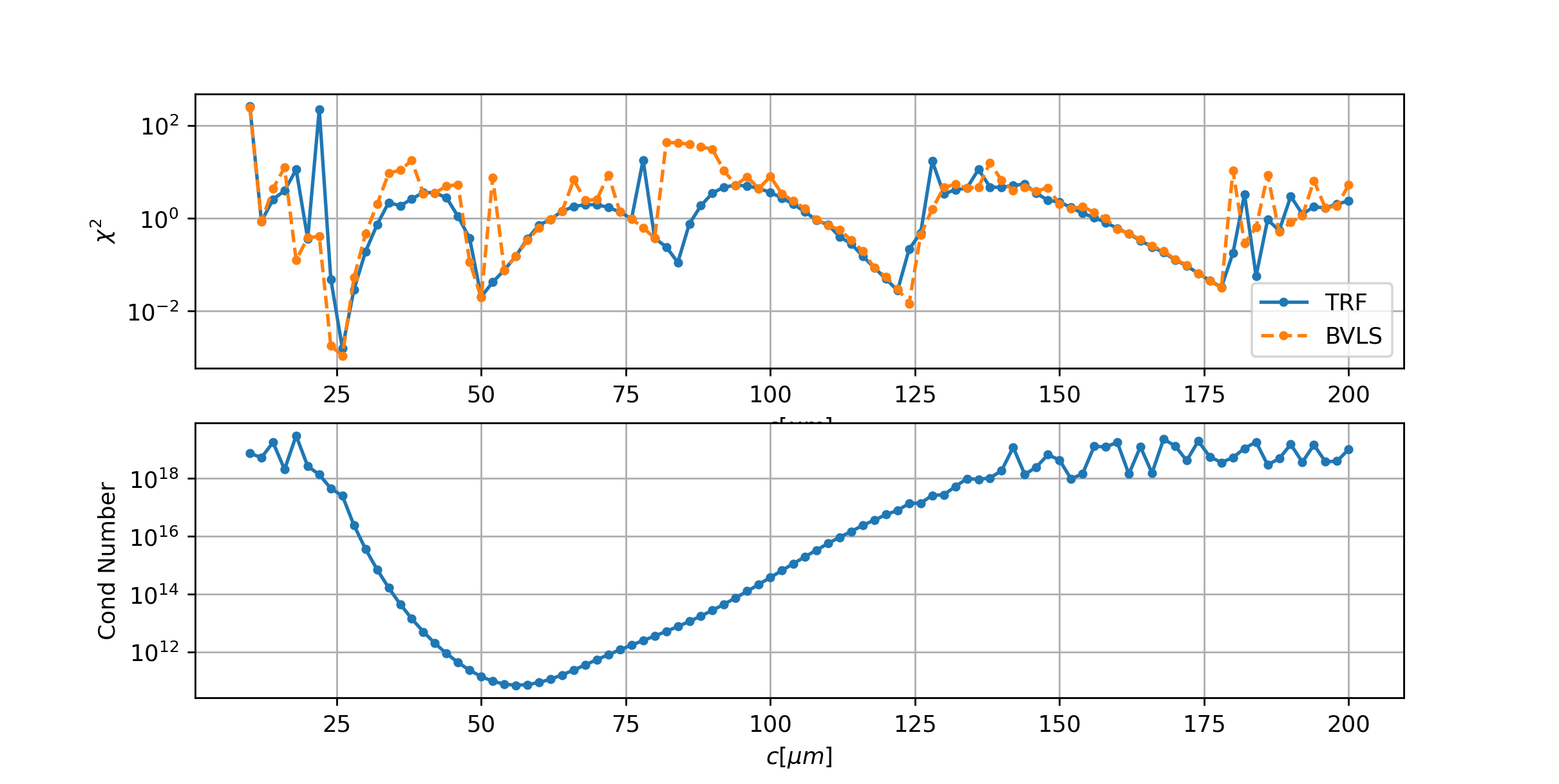}} \caption{Top: Evolution of $\chi^2$ vs the control electrode width $c$. The other parameters are given in the main text}
\label{fig:Chi2_1Case}
\end{figure}

For the particular case of 10 independent control electrodes ($n=10$), $h=100um$, $a=100\mu m$, $N = 64$, $d_{ion} = 10 \mu m$, $V_{DC}=0$, $V_{RF} = 200V$, $\Omega/2\pi = 20MHz$ and $V_{max} = 10V$, figure~\ref{fig:Chi2_1Case}a) shows the evolution of $\chi^2$ with the width of the control electrodes, $c$ (all taken to be equal, except the last outer ones, which are taken to be $c' = 4c$). The results of the two different algorithms are shown. The noise on the final residue is explained by the large values of the condition number of the $A$ matrix, shown in figure~\ref{fig:Chi2_1Case}b). In the following, both algorithms will be used and the one with the lower value of $\chi^2$ will be kept.

\begin{figure}
\center
\scalebox{0.5}{\includegraphics{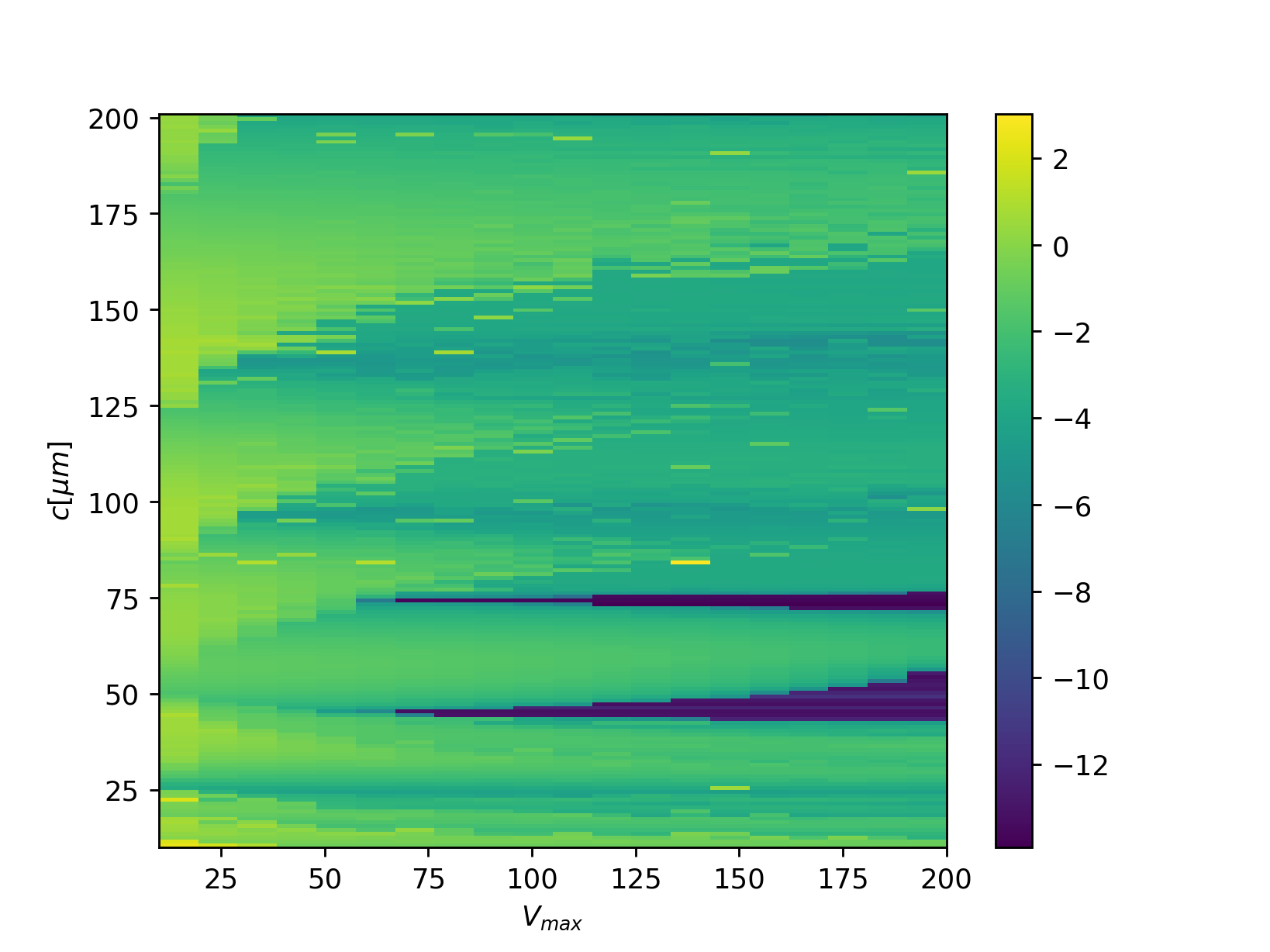}} \caption{Dependency of $\chi^2$ (in color-scale) vs the maximum voltage allowed, $V_{max}$, and the control electrodes width,$c$. The color axis correspond to $\log{\chi^2}$}.
\label{fig:vmax_vs_c}
\end{figure}

The dependency of $\chi^2$ with the electrode width, $c$ and the maximum allowed control voltage, $V_{max}$, is shown in figure~\ref{fig:vmax_vs_c}. The other parameters remain the same. We observe that by increasing the maximum voltage allowed, a region can be found  with a significantly lower value of $\chi^2$: from $\chi^2(V_{max} = 60, c=45um ) = 1.5\cdot10{-6}$ to $\chi^2(V_{max} = 70, c=45um ) = 6\cdot10{-14}$.

\begin{figure}
\center
\scalebox{0.5}{\includegraphics{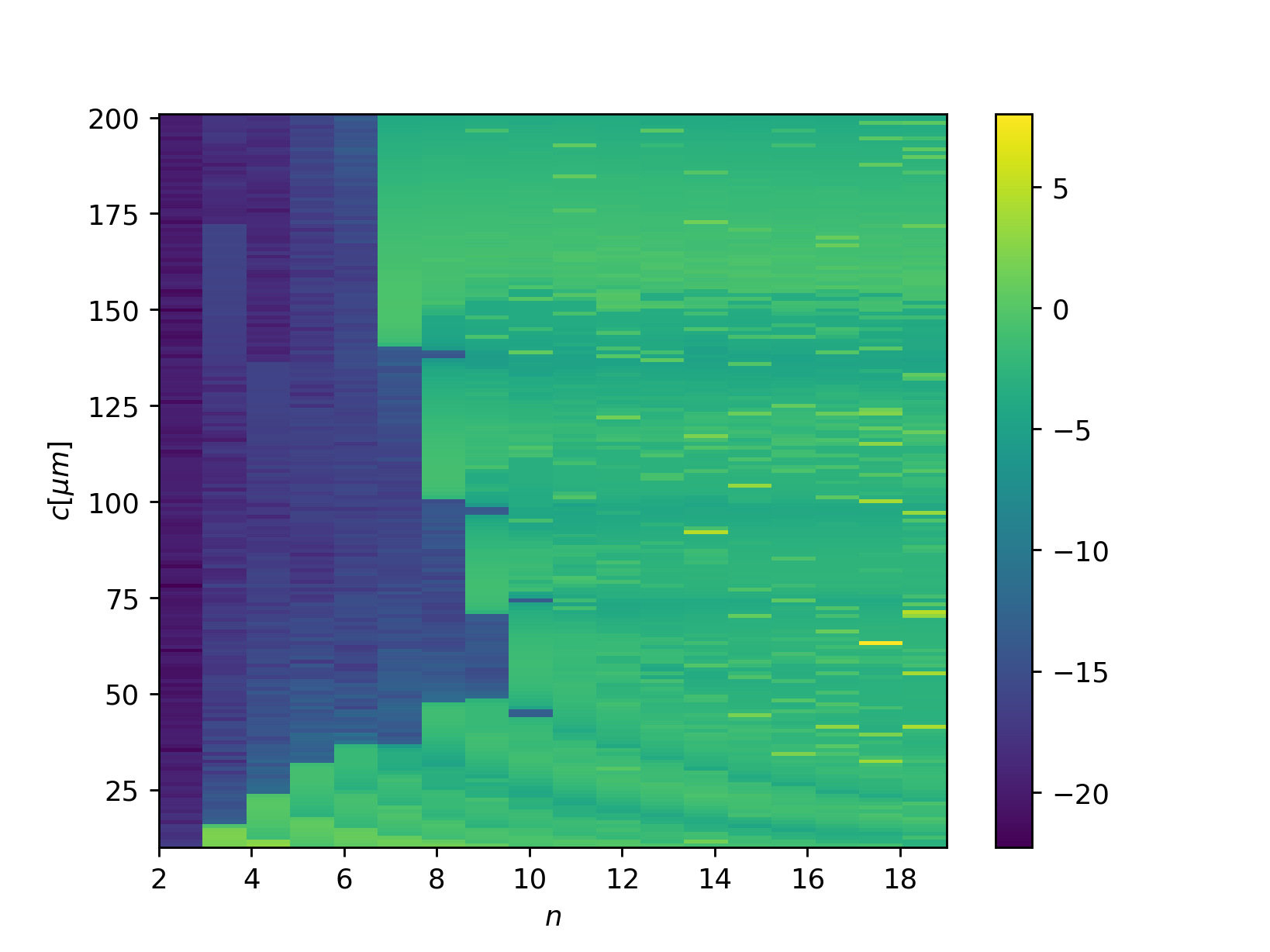}} \caption{Evolution of $\chi^2$ with the number of control electrodes $n$, and their width, $c$. The color axis correspond to $\log{\chi^2}$}
\label{fig:n_vs_c}
\end{figure}

Another variable that influences the uniformity, is the number of control electrodes for a fixed value of $V_{max}$. For the case of $V_{max} = 80$V,  we obtain figure~\ref{fig:n_vs_c}.

Notice that in figure~\ref{fig:n_vs_c}, there is a region with few number of control electrodes that leads to a very low value of $\chi^2$. However, the voltages provided by the optimization algorithm in such cases leads to axial potential depths (APD) that are also very low. For example, for $c = 45\mu m$ and $n=2$ and $n=4$, the axial potential depth is $V_{APD}\approx 3.5meV$, while for $n=10$ we obtain $V_{APD}\approx 70meV$. A higher value of $V_{APD}$ can be obtained by allowing a higher value $V_{max}$: for $V_{max}=100V$, $c=74\mu m$ and $n=10$, leads to $V_{APD} \approx 280meV$.



The results regarding the value of $\chi^2/N$ for a given set of trap and ion chain parameters provide useful information for the decision making during the design process. However, it does not directly provide information about the degree of homogeneity of the ion chain. Such issue is the topic of the next section.

\section{Molecular dynamic simulations}

In the following we will use the values of $V_{cc}$ obtained using the methodology presented above to perform realistic molecular dynamic simulations of the trapped ions. The equations of motion to be solved for each ion $i$, are given by:
\begin{align}
m \partial_{tt} x_i = Q^2 k_C\sum_{j=1,j\neq i}^N{\frac{x_i - x_{j}}{|\vec{r}_{i}-\vec{r}_{j}|^3}} + Q E^x(r_i)- \Gamma\partial_t x_i \nonumber\\
m \partial_{tt} y_i = Q^2 k_C\sum_{j=1,j\neq i}^N{\frac{y_i - y_{j}}{|\vec{r}_{i}-\vec{r}_{j}|^3}} + Q E^y(r_i)- \Gamma\partial_t y_i \nonumber \\
m \partial_{tt} z_i = Q^2 k_C\sum_{j=1,j\neq i}^N{\frac{z_i - z_{j}}{|\vec{r}_{i}-\vec{r}_{j}|^3}} + Q E^z(r_i)- \Gamma\partial_t z_i 
\end{align}
where $\Gamma$ is the friction coefficient due to laser cooling, $r = (x,y,z)$ and the electric field of trap is given by:

\begin{align}
E^j(r_i) & = V_{RF}\cos(\Omega t) E^{RF,j}_k(r_i) + V_{cc}E^{CC,j}_k(r_i)  \nonumber \\
         & + \sum_{k=1}^{n_c}{ V_k E^{CC,j}_k(r_i) } 
\end{align}
where $j$ denotes the orthogonal spatial components $(x,y,z)$. The C code which implements the GL approximation analytical solutions is a modification of a code reported in~\cite{pedregosa-gutierrez_direct_2021}.

\begin{figure}
\center
\scalebox{0.60}{\includegraphics{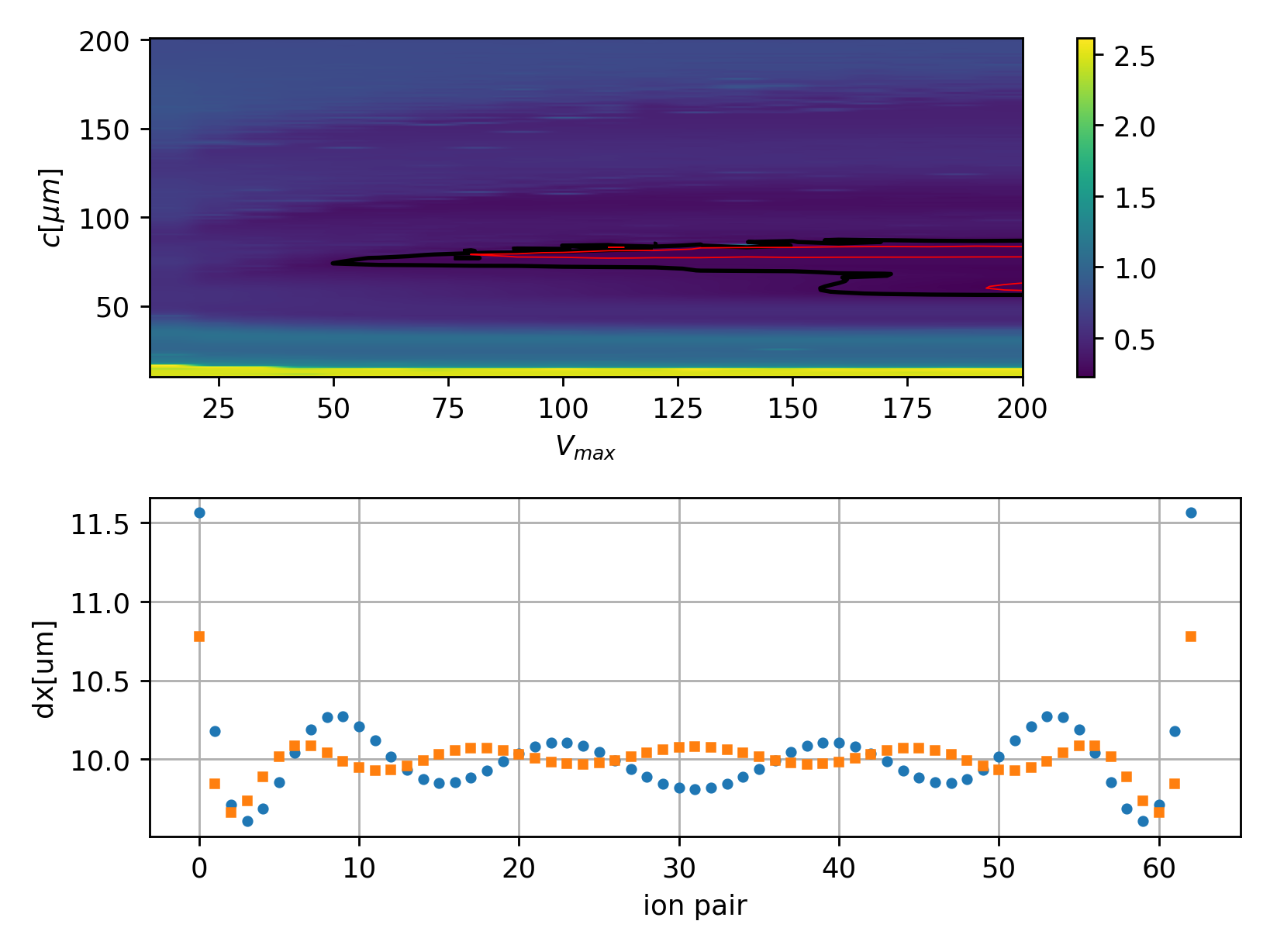}} \caption{Left: Dependency of inhomogeneity at the end of MD given by $\xi$, see text for details, versus the maximum voltage allowed, $V_{max}$ and the width of the control electrodes, $c$. The colour code indicates $\log{100\xi}$. The Black contour represents $\xi = 2\%$, while the Red contour represents the $\xi = 1.75\%$. Right: Two examples of final configurations. Big Blue Dots: $V_{max} = 10V$ and $c = 89\mu m$ leading to $\xi = 3.2\%$. Small Orange Dots: $V_{max} = 100V$ and $c = 80\mu m$ leading to $\xi = 1.6\%$.}
\label{fig:example_dd_N64}
\end{figure}

The ions are initialised at the desired position, i.e., as an homogenous ion chain with inter-ion distance equal $d$ at a height $h$. The trap frequency used is $\Omega / 2 \pi = 20MHz$, $V_{RF} = 200V$, $V_{DC} = 0V$, the friction term used is $\Gamma = 2.0\cdot 10^{-19} kg/s$ and the time-step of the Velocity-Verlet algorithm is $dt = \Omega / 100 = 2ns$. The final positions after $1~$ms simulation are used to evaluate the final inhomogeneity of the ions corresponding to a particular set of $V_{cc}$. The inhomogeneity is defined as the ratio between the spacing standard deviation and the mean of such spacing~\cite{pagano_cryogenic_2018}: $\xi = \sigma_{\Delta x} / \bar{\Delta x}$. By using the same parameter space as figure~\ref{fig:vmax_vs_c}, we obtain figure~\ref{fig:example_dd_N64}, where the colour code represents $\log_{10}(100\xi)$.

On the right side of figure~\ref{fig:example_dd_N64}, two examples of the ion-ion distances are shown. It is clear that the two extremes ions are responsible for a significant increase on the $\xi$ parameter. If such two ions are excluded on the computation of $\xi$, then the $\xi$ parameter would decrease from $3.2\% \to 1.5\%$ ($1.6\% \to 0.9\%$) for the $V_{max} = 10V$ ($V_{max} = 100V$) case shown. 

\begin{figure}
\center
\scalebox{0.5}{\includegraphics{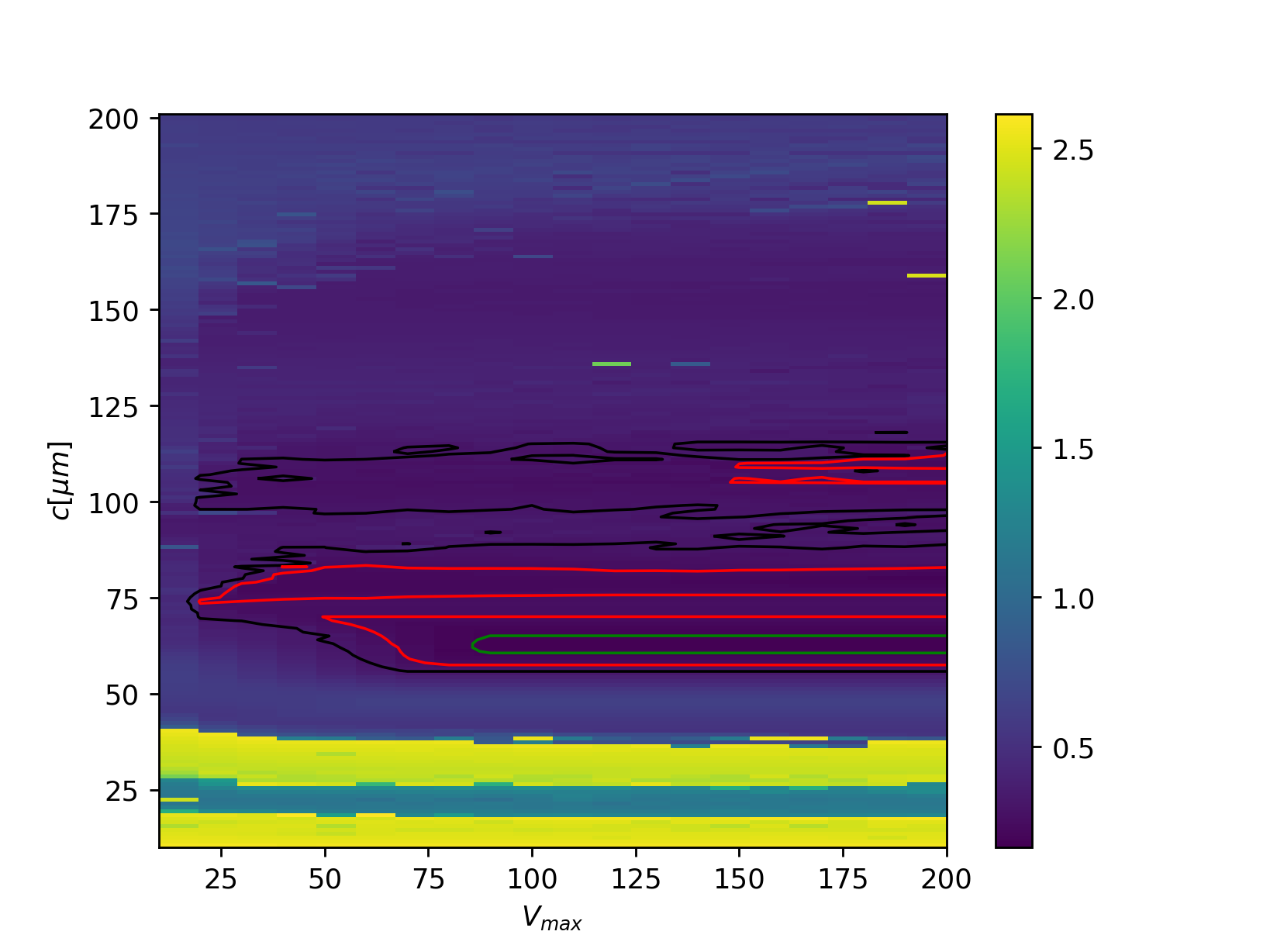}} \caption{Left: Similar to figure~\ref{fig:example_dd_N64} where the two extreme ions have been excluded of the optimization method to obtain the set of $V_{cc}$ voltages.}
\label{fig:N64_2D_Without_Extremes}
\end{figure}

In fact, the two extreme ions can be removed from the optimization problem, i.e., we find the set of $V_{cc}$ that leads to the more homogeneous ion chain, with the exception of the two extreme ions. If we perform again molecular dynamic simulations with such new sets of $V_{cc}$, it leads to figure~\ref{fig:N64_2D_Without_Extremes}. The range of parameters where the degree of homogeneity is lower than 2\% has significantly increased and it is now possible to achieve values of $\xi<1.5$ (green contour). The approach can be extended to exclude a greater number of ions from each side, which it will improve even further the degree of homogeneity. Such flexibility could be exploited to compensate for technical limitations on the possible maximum values or sizes of the electrodes.

\section{From Harmonic trap to homogeneous ion chain}

The set of voltages leading to an homogeneous ion chain are not necessarily the more appropriate choice to trap and cool the ion chain down to ground state. For this reason, we perform molecular dynamic simulations where the ions are initialised in the well known and studied~\cite{RN133} harmonic potential, followed by a transition to the values corresponding to the homogeneous ion chain.

The half-length of an ion chain in an harmonic potential of secular frequency, $\omega$, can be estimated to be~\cite{dubin_minimum_1997}:
\begin{equation}
L^3= \frac{3 N Q^2 k_C}{m \omega^2} (\ln N + \ln + \gamma_e -7/2 )
\end{equation}

We have use an approach similar as above to find the needed voltages to apply in order to obtain an harmonic potential with $\omega / 2\pi = 212kHz$, which should lead to a total ion chain length similar for both the harmonic potential and the homogeneous potential case.

\begin{figure}
\center
\scalebox{0.35}{\includegraphics{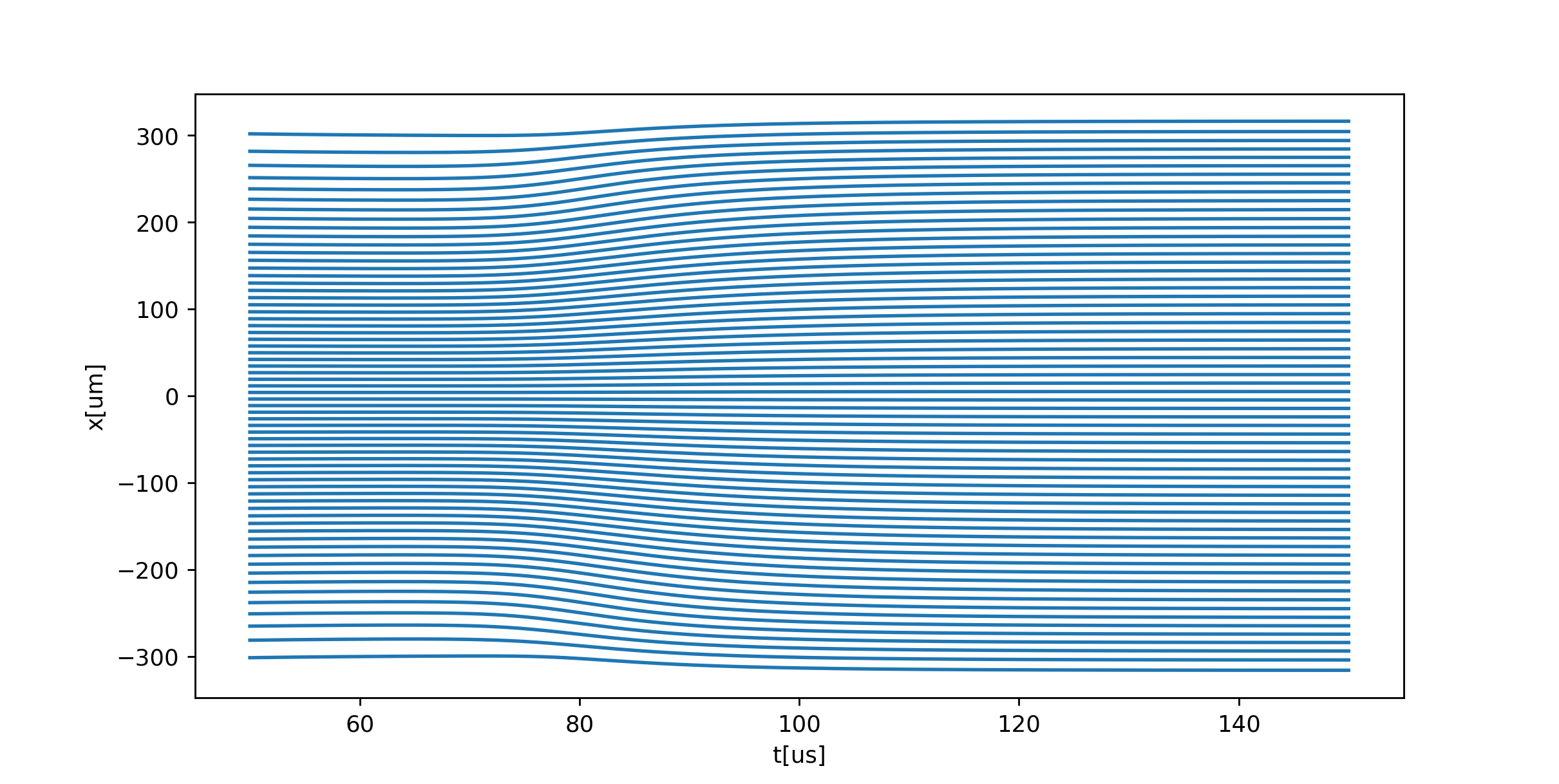}} \caption{Evolution of the axial position when the trapping voltages are evolved using a "tangent gate" (see main text for details) from an harmonic potential to the homogeneous one.}
\label{fig:From_Harmo_to_Homo}
\end{figure}

After initializing the ions in the harmonic potential, the simulation waits for $50\mu$s before the transition starts to the new set of voltage values. A transition time of $t_g = 50\mu$s has been used. After the transition has finished, the simulation continues during $50\mu$s more. In order to change smoothly the voltages from the harmonic to the homogenous case, we have used an hyperbolic tangent transition, given by~\cite{pedregosa-gutierrez_ion_2015,kamsap_fast_2015, kamsap_fast_2015-1}:
\begin{equation}
    V(t) = V_{1} + \frac{V_{2} - V_{1} }{2} \left( \frac{\tanh(n_h(2t/t_g - 1))}{\tanh(n_h)}  + 1\right)
\end{equation}
where $V_1$ correspond to the set of voltages needed for the harmonic case, $V_2$ correspond to the set of voltages needed for the homogeneous case and $n_h$ parameter is used to go from a linear evolution ($n_h = 1$) to a step function ($n_h \to \infty$). The evolution of the axial position of the ions is shown in figure~\ref{fig:From_Harmo_to_Homo}, where $n_h = 4$ has been used. It illustrates that the approach proposed should be feasible and applicable to existing surface traps.

\section{Conclusion}
In summary, we have developed a novel algorithm capable to obtain the control electrodes voltages of a surface ion trap needed to generate, with high fidelity, an uniform ion distribution and an harmonic ion distribution. The approach followed is general and therefore, it should be applicable to a large range of desired ion distributions. The approach is based on local minimization of the total force field at each individual ion position. While the present work has used the obtained the needed electric-fields using the GL approximation, the method can be use electric-fields obtained by FEM/BEM methods to more accurately represent the final geometry.

The validity of the different voltages sets obtained by our method have been  cross-checked against a robust molecular dynamics simulation. The simulation results show that even for a relatively large ion number, uniformity below $2\%$ is achievable with only $10$ pairs of electrodes and limited voltage applied to them. Therefore, it opens up the possibilities to implement an uniform chain of trapped and laser cooled ions for realizing homogeneous KZM, implementing large scale entanglement using the BHM and scaling up the nodes of a distributed quantum computing architecture.


\begin{thebibliography}{10}

\bibitem{bruzewicz_trapped-ion_2019}
Colin~D. Bruzewicz, John Chiaverini, Robert McConnell, and Jeremy~M. Sage.
\newblock Trapped-ion quantum computing: {Progress} and challenges.
\newblock {\em Applied Physics Reviews}, 6(2):021314, June 2019.
\newblock Publisher: American Institute of Physics.

\bibitem{PhysRevLett.97.243005}
A.~Ostendorf, C.~B. Zhang, M.~A. Wilson, D.~Offenberg, B.~Roth, and
  S.~Schiller.
\newblock Sympathetic cooling of complex molecular ions to millikelvin
  temperatures.
\newblock {\em Phys. Rev. Lett.}, 97:243005, Dec 2006.

\bibitem{RN174}
L.~Schmöger, O.~O. Versolato, M.~Schwarz, M.~Kohnen, A.~Windberger, B.~Piest,
  S.~Feuchtenbeiner, J.~Pedregosa-Gutierrez, T.~Leopold, P.~Micke, A.~K.
  Hansen, T.~M. Baumann, M.~Drewsen, J.~Ullrich, P.~O. Schmidt, and
  J.~R.~Crespo López-Urrutia.
\newblock Coulomb crystallization of highly charged ions.
\newblock {\em Science}, 347(6227):1233, 2015.

\bibitem{march_practical_1995}
Raymond~E. March and John F.~J. Todd, editors.
\newblock {\em Practical aspects of ion trap mass spectrometry}.
\newblock Modern mass spectrometry. CRC Press, Boca Raton, Fla, 1995.

\bibitem{pino_demonstration_2021}
J.~M. Pino, J.~M. Dreiling, C.~Figgatt, J.~P. Gaebler, S.~A. Moses, M.~S.
  Allman, C.~H. Baldwin, M.~Foss-Feig, D.~Hayes, K.~Mayer, C.~Ryan-Anderson,
  and B.~Neyenhuis.
\newblock Demonstration of the trapped-ion quantum {CCD} computer architecture.
\newblock {\em Nature}, 592(7853):209--213, April 2021.
\newblock Number: 7853 Publisher: Nature Publishing Group.

\bibitem{hucul_transport_2008}
D.~Hucul, M.~Yeo, W.~K. Hensinger, J.~Rabchuk, S.~Olmschenk, and C.~Monroe.
\newblock On the {Transport} of {Atomic} {Ions} in {Linear} and
  {Multidimensional} {Ion} {Trap} {Arrays}.
\newblock {\em arXiv:quant-ph/0702175}, February 2008.
\newblock arXiv: quant-ph/0702175.

\bibitem{amini_toward_2010}
J~M Amini, H~Uys, J~H Wesenberg, S~Seidelin, J~Britton, J~J Bollinger,
  D~Leibfried, C~Ospelkaus, A~P VanDevender, and D~J Wineland.
\newblock Toward scalable ion traps for quantum information processing.
\newblock {\em New Journal of Physics}, 12(3):033031, March 2010.

\bibitem{mokhberi_optimised_2017}
A~Mokhberi, R~Schmied, and S~Willitsch.
\newblock Optimised surface-electrode ion-trap junctions for experiments with
  cold molecular ions.
\newblock {\em New Journal of Physics}, 19(4):043023, April 2017.

\bibitem{RN134}
D.~Kielpinski, C.~Monroe, and D.~J. Wineland.
\newblock Architecture for a large-scale ion-trap quantum computer.
\newblock {\em Nature}, 417(6890):709--711, 2002.

\bibitem{warschburger_experimental_2014}
C.~Warschburger, H.~Kaufmann, C.~T. Schmiegelow, A.~Walther, M.~Hettrich,
  A.~Pfister, V.~Kaushal, F.~Schmidt-Kaler, U.~G. Poschinger, and T.~Ruster.
\newblock Experimental realization of fast ion separation in segmented {Paul}
  traps.
\newblock {\em Physical Review A}, 90(3):033410, September 2014.
\newblock Publisher: American Physical Society.

\bibitem{RN135}
C.~Monroe and J.~Kim.
\newblock Scaling the ion trap quantum processor.
\newblock {\em Science}, 339(6124):1164, 2013.

\bibitem{kamsap_experimental_2017}
M.~R. Kamsap, C.~Champenois, J.~Pedregosa-Gutierrez, S.~Mahler, M.~Houssin, and
  M.~Knoop.
\newblock Experimental demonstration of an efficient number diagnostic for long
  ion chains.
\newblock {\em Physical Review A}, 95(1):013413, January 2017.

\bibitem{RN161}
J.~Pedregosa-Gutierrez and M.~Mukherjee.
\newblock Defect generation and dynamics during quenching in finite size
  homogeneous ion chains.
\newblock {\em New Journal of Physics}, 22(7):073044, 2020.

\bibitem{RN73}
T.~Dutta, M.~Mukherjee, and K.~Sengupta.
\newblock Ramp dynamics of phonons in an ion trap: Entanglement generation and
  cooling.
\newblock {\em Physical Review Letters}, 111(17):5, 2013.

\bibitem{lin_large-scale_2009}
G.-D. Lin, S.-L. Zhu, R.~Islam, K.~Kim, M.-S. Chang, S.~Korenblit, C.~Monroe,
  and L.-M. Duan.
\newblock Large-scale quantum computation in an anharmonic linear ion trap.
\newblock {\em EPL (Europhysics Letters)}, 86(6):60004, June 2009.

\bibitem{xie_creating_2017}
Yi~Xie, Xinfang Zhang, Baoquan Ou, Ting Chen, Jie Zhang, Chunwang Wu, Wei Wu,
  and Pingxing Chen.
\newblock Creating equally spaced linear ion string in a surface-electrode trap
  by feedback control.
\newblock {\em Physical Review A}, 95(3):032341, March 2017.

\bibitem{house_analytic_2008}
M.~G. House.
\newblock Analytic model for electrostatic fields in surface-electrode ion
  traps.
\newblock {\em Physical Review A}, 78(3):033402, September 2008.

\bibitem{schmied_electrostatics_2010}
Roman Schmied.
\newblock Electrostatics of gapped and finite surface electrodes.
\newblock {\em New Journal of Physics}, 12(2):023038, February 2010.

\bibitem{turchette_heating_2000}
Q.~A. Turchette, {Kielpinski}, B.~E. King, D.~Leibfried, D.~M. Meekhof, C.~J.
  Myatt, M.~A. Rowe, C.~A. Sackett, C.~S. Wood, W.~M. Itano, C.~Monroe, and
  D.~J. Wineland.
\newblock Heating of trapped ions from the quantum ground state.
\newblock {\em Physical Review A}, 61(6):063418, May 2000.

\bibitem{wineland_experimental_1998}
D.~J. Wineland, C.~Monroe, W.~M. Itano, D.~Leibfried, B.~E. King, and D.~M.
  Meekhof.
\newblock Experimental {Issues} in {Coherent} {Quantum}-{State} {Manipulation}
  of {Trapped} {Atomic} {Ions}.
\newblock {\em Journal of Research of the National Institute of Standards and
  Technology}, 103(3):259--328, 1998.

\bibitem{bevington_data_2003}
Philip~R. Bevington and D.~Keith Robinson.
\newblock {\em Data reduction and error analysis for the physical sciences}.
\newblock McGraw-Hill, Boston, 3rd ed edition, 2003.

\bibitem{virtanen_scipy_2020}
Pauli Virtanen, Ralf Gommers, Travis~E. Oliphant, Matt Haberland, Tyler Reddy,
  David Cournapeau, Evgeni Burovski, Pearu Peterson, Warren Weckesser, Jonathan
  Bright, Stéfan~J. van~der Walt, Matthew Brett, Joshua Wilson, K.~Jarrod
  Millman, Nikolay Mayorov, Andrew R.~J. Nelson, Eric Jones, Robert Kern, Eric
  Larson, C.~J. Carey, İlhan Polat, Yu~Feng, Eric~W. Moore, Jake VanderPlas,
  Denis Laxalde, Josef Perktold, Robert Cimrman, Ian Henriksen, E.~A. Quintero,
  Charles~R. Harris, Anne~M. Archibald, Antônio~H. Ribeiro, Fabian Pedregosa,
  and Paul van Mulbregt.
\newblock {SciPy} 1.0: fundamental algorithms for scientific computing in
  {Python}.
\newblock {\em Nature Methods}, 17(3):261--272, March 2020.
\newblock Number: 3 Publisher: Nature Publishing Group.

\bibitem{branch_subspace_1999}
Mary~Ann Branch, Thomas~F. Coleman, and Yuying Li.
\newblock A {Subspace}, {Interior}, and {Conjugate} {Gradient} {Method} for
  {Large}-{Scale} {Bound}-{Constrained} {Minimization} {Problems}.
\newblock {\em SIAM Journal on Scientific Computing}, 21(1):1--23, January
  1999.
\newblock Publisher: Society for Industrial and Applied Mathematics.

\bibitem{stark_bounded-variable_1995}
P.~B. Stark and R.~L. Parker.
\newblock Bounded-variable least-squares: an algorithm and applications.
\newblock {\em Computational Statistics}, 1995.

\bibitem{pedregosa-gutierrez_direct_2021}
Jofre Pedregosa-Gutierrez and Jim Dempsey.
\newblock Direct {N}-{Body} problem optimisation using the {AVX}-512
  instruction set.
\newblock {\em arXiv:2106.11143 [physics]}, June 2021.
\newblock arXiv: 2106.11143.

\bibitem{pagano_cryogenic_2018}
G~Pagano, P~W Hess, H~B Kaplan, W~L Tan, P~Richerme, P~Becker, A~Kyprianidis,
  J~Zhang, E~Birckelbaw, M~R Hernandez, Y~Wu, and C~Monroe.
\newblock Cryogenic trapped-ion system for large scale quantum simulation.
\newblock {\em Quantum Science and Technology}, 4(1):014004, October 2018.

\bibitem{RN133}
D.~Leibfried, R.~Blatt, C.~Monroe, and D.~Wineland.
\newblock Quantum dynamics of single trapped ions.
\newblock {\em Reviews of Modern Physics}, 75(1):281--324, 2003.

\bibitem{dubin_minimum_1997}
Daniel H.~E. Dubin.
\newblock Minimum energy state of the one-dimensional {Coulomb} chain.
\newblock {\em Physical Review E}, 55(4):4017--4028, April 1997.

\bibitem{pedregosa-gutierrez_ion_2015}
Jofre Pedregosa-Gutierrez, Caroline Champenois, Marius~Romuald Kamsap, and
  Martina Knoop.
\newblock Ion transport in macroscopic {RF} linear traps.
\newblock {\em International Journal of Mass Spectrometry}, 381-382:33--40, May
  2015.

\bibitem{kamsap_fast_2015}
M.~R. Kamsap, C.~Champenois, J.~Pedregosa-Gutierrez, M.~Houssin, and M.~Knoop.
\newblock Fast accumulation of ions in a dual trap.
\newblock {\em EPL (Europhysics Letters)}, 110(6):63002, June 2015.

\bibitem{kamsap_fast_2015-1}
M.~R. Kamsap, J.~Pedregosa-Gutierrez, C.~Champenois, D.~Guyomarc'h, M.~Houssin,
  and M.~Knoop.
\newblock Fast and efficient transport of large ion clouds.
\newblock {\em Physical Review A}, 92(4):043416, October 2015.

\end{thebibliography}

\end{document}